# A two-stage diffusion-driven morphogenesis model with evolving connectivity

Alexander Safonov[a*]

[a]: Skolkovo Institute of Science and Technology, Center for Materials Technologies, Bolshoy Boulevard 30, bld. 1 Moscow, 121205, Russia

Corresponding Author: Alexander A. Safonov, PhD

Corresponding Author's Institution: Skolkovo Institute of Science and Technology

*Corresponding Author E-mail: a.safonov@skoltech.ru

**Abstract**

Morphogenesis of diatom silica shells involves a sequence of topological transitions: initially separate ribs grow and branch, then merge through bridges, leaving a regular network of pores. Existing models based on diffusion-limited aggregation or reaction-diffusion reproduce either branching or pore formation, but not both within a single framework. Here we propose a two-stage model of diffusion-driven morphogenesis with evolving connectivity. The model is formulated on a fixed finite element mesh with a density field on an active layer. Stage 1 reproduces dendritic growth, where a stimulus based on the local diffusive flux exceeds a threshold. Stage 2 uses an inverted stimulus, where high local flux suppresses precipitation, stabilizing pores between the dendrites. The update rules are analogous to sensitivity in density-based topology optimization and to strain energy density in bone remodeling. The model predicts distinct morphological regimes on each stage, controlled by the respective thresholds. For a reference calculation, the predicted pore network is regular, with approximately uniform spacing, and pores are absent near the domain edges, consistent with observations on diatom valves. The model provides a unified framework for describing both stages of silica morphogenesis and suggests a route for comparing computational predictions with cryo-electron tomography data.

## 1. Introduction

Morphogenesis, the emergence of pattern and form from initially uniform states, is a phenomenon studied across biology, geomorphology, and materials science. Since Turing's 1952 work on the chemical basis of pattern formation, it has been recognized that similar reaction-diffusion mechanisms operate in diverse natural systems [1,2]. Modeling studies of morphogenesis have since become a central tool for understanding how complex shapes emerge from local rules, spanning scales from molecular self-assembly to tissue mechanics [3]. The same general principles that produce sand ripples and zebra stripes may underlie the formation of biological structures such as diatom silica shells. Similar mechanisms have been shown to govern landscape evolution, including the formation of arcades and rock pillars through erosion [4] and honeycomb weathering through moisture transport and salt deposition [5]. Computational morphogenesis, the iterative redistribution of material to generate structural form, has been demonstrated at giga-voxel resolution in engineering design [6], where the resulting structures show similarity to natural bone. Computational morphogenesis has also been linked to natural shape optimization [7], to biological computing systems such as *Physarum polycephalum* [8], and to material topology optimization as a computational process [9]. Related methods have been extended to three-dimensional problems with anisotropic materials [10]. A dynamical-systems view of topology optimization and bone remodeling has been developed by Klarbring and Torstenfelt [11,12], who showed that update rules of density-based topology optimization can be interpreted as discretizations of evolution equations, and that natural remodeling processes share the same mathematical structure. Recent advances in density-based methods have extended topology optimization to multi-material and dynamic problems [13], and phase-field approaches combined with optimality criteria have been developed for topology optimization [14]. This work continues a line of research on natural morphogenesis as a computational process [15].

Diatoms are unicellular algae that produce ornate, amorphous silica cell walls with structures spanning several orders of magnitude in size [16]. With an estimated 100,000 species, diatoms display a wide diversity of morphologies, yet the mechanisms controlling silica patterning within the cell remain unresolved [17]. The process occurs inside a membrane-bound organelle, the silica deposition vesicle (SDV). Recent cryo-electron tomography has revealed that the SDV membrane acts as a dynamic mold that shapes the forming silica [17]. The curvature of the tethered side of the SDV membrane mirrors the silica topography, suggesting that silica growth and morphogenesis result from the biophysical properties of the SDV and plasma membranes. This provides a biological motivation for computational models where local membrane properties control the boundary conditions for diffusion and precipitation.

Diatom silica morphology exhibits evolving connectivity: the shell is not a static pattern but a structure whose topology changes during formation. Initially separate ribs grow and branch, then merge through bridges, leaving a regular network of pores. Figure 1 provides a direct illustration of this two-stage process: panel (a) shows the sequential development of ribs and cross-bridges in *Navicula pelliculosa* [18], while panel (b) shows the resulting regular pore network in *Stephanodiscus hantzschii* [19]. Existing models capture only one aspect of this process. Diffusion-limited aggregation models reproduce dendritic branching, but do not account for the subsequent formation of pores [20]. Recent reaction-diffusion models quantitatively recapitulate the branching of silica ribs, but do not describe pore formation [21]. Cellular-automaton models have reproduced pore patterns in pennate diatoms, but not the branching of centric species [22]. Conversely, phase-separation models describe the formation of regular pore patterns, but assume the pre-existence of the rib scaffold [23,24]. A unified model in which both stages are described within a single framework, and in which the transition between them is an emergent property rather than an imposed condition, is still lacking.

Here we propose a two-stage model of diffusion-driven morphogenesis with evolving connectivity. The model is formulated on a fixed finite element mesh with a density field $\rho$ on an active layer. Stage 1 reproduces dendritic growth, where a stimulus based on the local diffusive flux exceeds a threshold. Stage 2 uses an inverted stimulus, where high local flux suppresses precipitation, stabilizing pores between the dendrites. The model predicts four regimes on each stage, controlled by the respective thresholds. The combination defines a phase diagram of morphologies. The present work focuses on the methodological framework and its qualitative predictions. Application to specific biological systems, including comparison with cryo-ET data, is a subject of ongoing work.

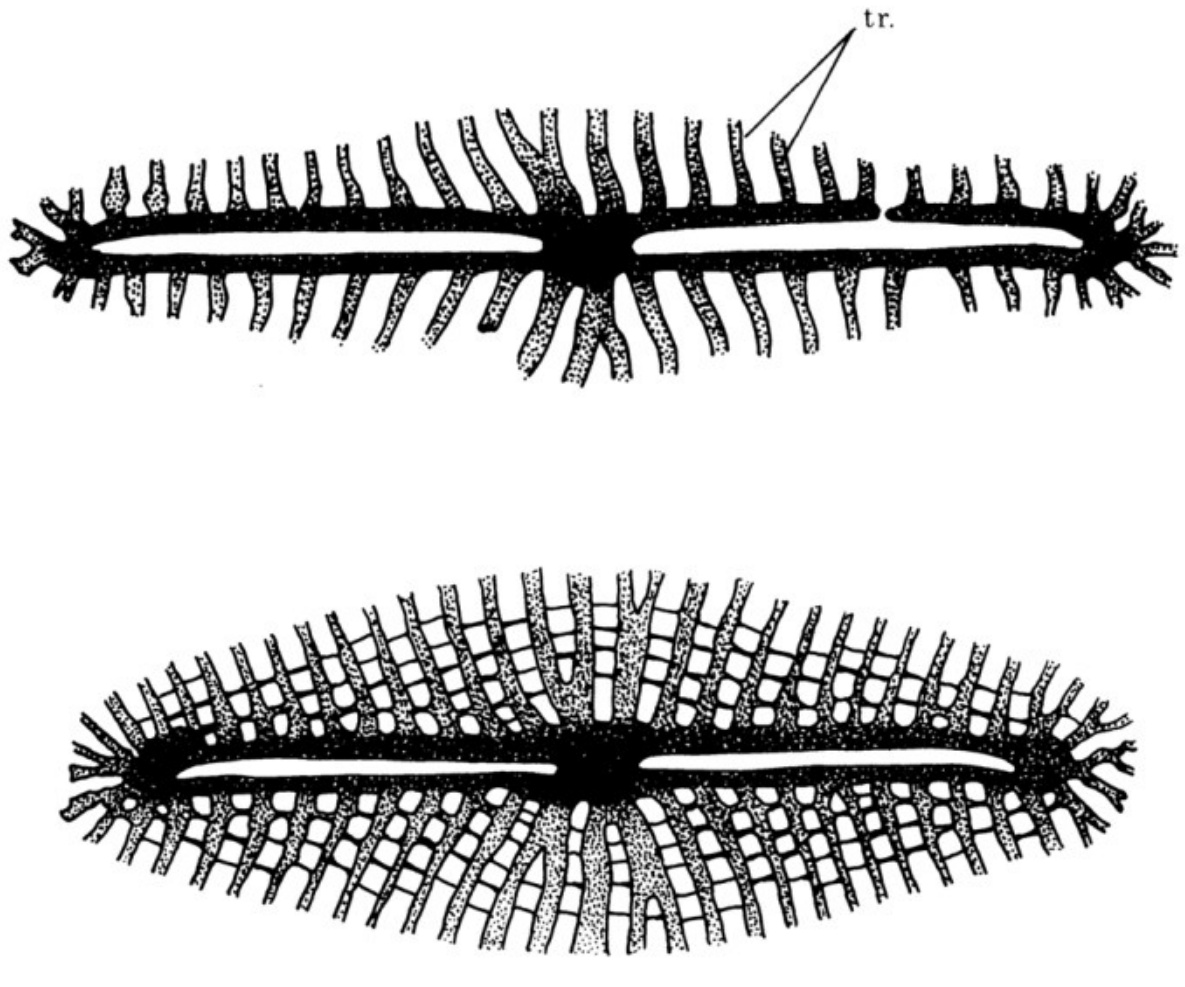

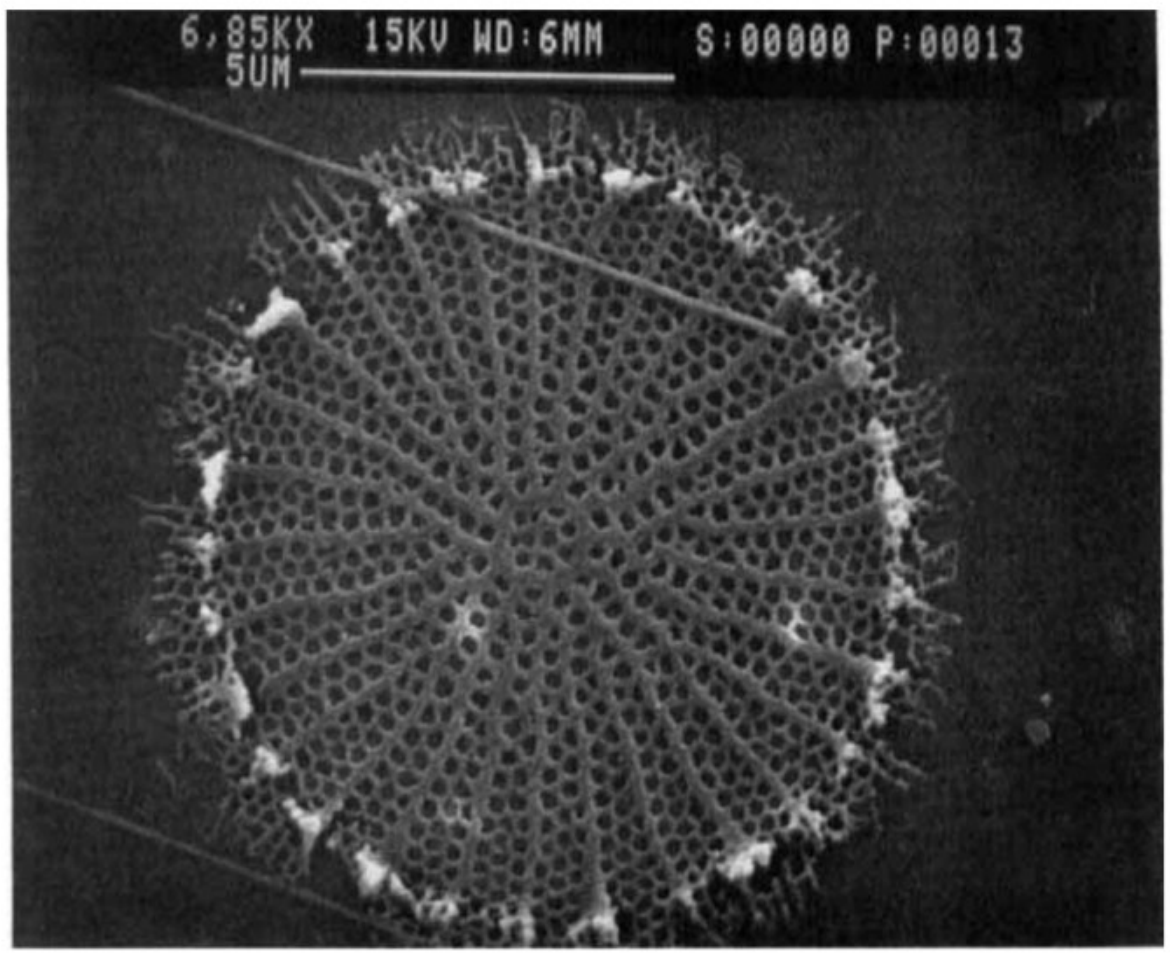

**Figure 1.** Biological motivation for the two-stage model. (a) Sequential valve development in the pennate diatom *Navicula pelliculosa*: transapical ribs grow first (top), followed by the formation of cross-bridges that delimit pores (bottom). Adapted from [18]. (b) SEM image of a partially dissolved valve of the centric diatom *Stephanodiscus hantzschii*, showing a regular hexagonal pore network. Adapted from [19]. The present model reproduces this sequence: Stage 1 corresponds to rib/dendrite growth, Stage 2 to bridge formation and pore creation.

## 2. Model

### 2.1. Model overview

The model describes the formation of a silica structure on an active layer via a two-stage iterative scheme. The primary variable is the concentration field $u(\mathbf{x})$, which represents the concentration of dissolved matter within the surrounding volume. The local flux is $\mathbf{J} = -D\nabla u$, with $D$ the diffusion coefficient.

The model proceeds in two stages. In **Stage 1** (dendritic growth), the flux supplies matter to the growing structures and promotes precipitation. In **Stage 2** (pore and bridge formation), the flux removes matter from the inter-dendritic regions and suppresses precipitation. The flux **J** is treated as a **generic matter flux**, without distinguishing its physical components; the distinction of these components and their simultaneous action is left for future work.

At each iteration, the stationary diffusion problem is solved:

$$\Delta u = 0 \ \text{ in } \Omega.$$

The model is formulated on a rectangular parallelepiped $\Omega \subset \mathbb{R}^3$ discretized by a regular finite element mesh of cubic elements. The domain has six faces (Fig. 2a):

- **Source face** ($\Gamma_s$): fixed concentration $u = u_0$.
- **Active layer** ($\Gamma_a$): one element thick, where silica precipitation occurs. The active layer is in contact with a reservoir of fixed concentration $u_s$, representing the sink for dissolved matter. The properties of this contact are described in Section 2.2.
- **Symmetry faces** ($\Gamma_{\text{sym}}$): $\partial_n u = 0$ (two opposite faces), allowing the computational domain to represent half of a symmetric system.
- **Edge faces** ($\Gamma_{\text{edge}}$): fixed concentration $u = u_0$, representing the SDV boundary.

The stimulus is defined in terms of $|J|^2 = J_x^2 + J_y^2 + J_z^2$, evaluated at the integration points of the active layer and averaged over each element.

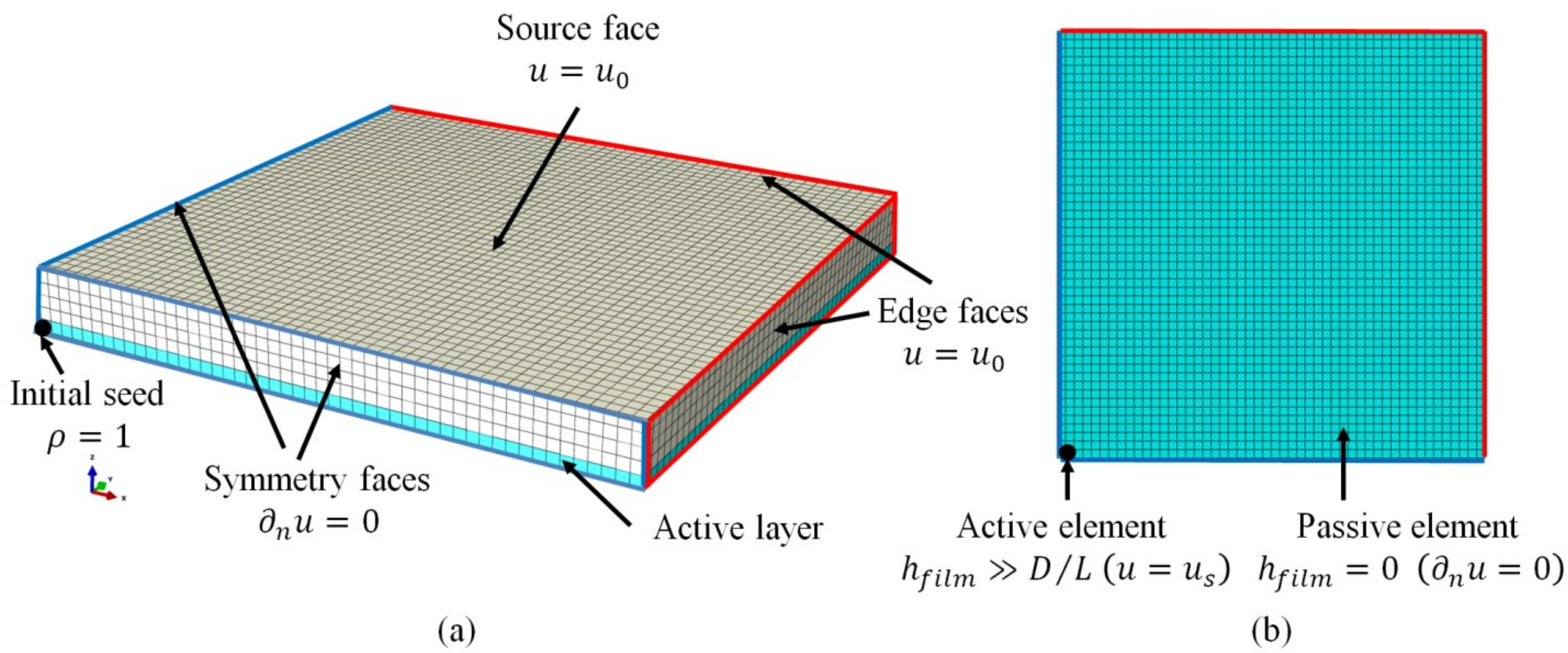

**Figure 2**. Model setup. (a) Computational domain (quarter-model) with boundary conditions. (b) Boundary condition on the active layer: Robin condition with film coefficient $h_{\text{film}}$ for active elements and Neumann for passive elements.

### 2.2. Density field and analogy with density-based topology optimization

A scalar density field $\rho(\mathbf{x}) \in [\rho_{\min}, 1]$ is defined on the active layer, representing the local volume fraction of precipitated silica. $\rho = 1$ corresponds to fully dense silica, while $\rho = \rho_{\min}$ represents an empty (pore) region. This is analogous to the density field in the SIMP (Solid Isotropic Material with Penalization) method for topology optimization, where $\rho$ interpolates between solid material and void [25], [26]. The lower bound $\rho_{\min}$ prevents singularity of the equilibrium problem and is standard in density-based methods [27].

However, unlike in SIMP, the diffusion coefficient $D$ does not depend on $\rho$. The density affects the boundary conditions on the active layer and the update rule, but not the bulk diffusion. This is justified by the observation that the main diffusion resistance is in the inactive layers, while the active layer is thin and its contribution to the overall transport is negligible.

The density controls the boundary condition on the active layer through a Robin condition applied to all elements (Fig. 2b):

$$-D\partial_n u = h_{\text{film}}(u - u_s),$$

where $u_s$ is the concentration in the adjacent reservoir and the film coefficient $h_{\mathrm{film}}$ depends on the state of the element. Elements are either **active** (acting as sinks) or **passive**. Active elements are assigned a large film coefficient ($h_{\mathrm{film}} \gg D/L$), while passive elements have $h_{\mathrm{film}} = 0$, so that the condition reduces to Neumann ($\partial_n u = 0$). When $h_{\mathrm{film}} L/D \gg 1$, the active condition is numerically equivalent to Dirichlet $u = u_s$.

Two types of thresholds are used in the model. The **activation thresholds** $\rho_{\mathrm{act}}^{(1)}$ and $\rho_{\mathrm{act}}^{(2)}$ control when an element becomes active (a sink): elements with $\rho \geq \rho_{\mathrm{act}}^{(1)}$ are active in Stage 1, and elements with $\rho \leq \rho_{\mathrm{act}}^{(2)}$ are active in Stage 2. The **dissolution threshold** $\rho_{\mathrm{sol}}$ controls the irreversibility of dendritic growth: elements with $\rho \geq \rho_{\mathrm{sol}}$ cannot decrease in density, representing the fact that precipitated silica does not redissolve.

### 2.3. Two-stage iterative scheme

In **Stage 1** (dendritic growth), elements with $\rho \geq \rho_{\mathrm{act}}^{(1)}$ are active. The stimulus is

$$S_1 = \frac{|J|^2}{\rho}.$$

In this stage, the flux is interpreted as the diffusion of dissolved silica, which links the present model to earlier works on silica transport and precipitation in diatoms [28,29]. The form of $S_1$ is analogous to the strain energy density per unit mass used in bone remodeling simulations [30] and to the sensitivity of compliance in SIMP-based topology optimization, where the design sensitivity scales as $|J|^2/\rho$ [25]. The density is updated according to

$$\rho^{(n+1)} = \rho^{(n)} + \delta\rho \cdot \begin{cases} +1, & S_1 \geq \theta_1 \\ 0, & S_1 < \theta_1, \rho \geq \rho_{\mathrm{sol}} \\ -1/10, & S_1 < \theta_1, \rho < \rho_{\mathrm{sol}} \end{cases}$$

where $\delta\rho$ is the density update step and $\theta_1$ is the threshold. The density is constrained to remain within $[\rho_{\min}, 1]$. Stage 1 runs for a prescribed number of iterations $n_1$.

In **Stage 2** (pore and bridge formation), elements with $\rho \leq \rho_{\mathrm{act}}^{(2)}$ become active (pores). In the stimulus, the dependence on density is inverted, while the flux term remains unchanged:

$$S_2 = \frac{|J|^2}{1-\rho+\varepsilon},$$

with $\varepsilon$ a small regularization parameter. In this stage, the flux is interpreted as the transport of matter through the pores, which suppresses further precipitation. The update rule is inverted with respect to Stage 1:

$$\rho^{(n+1)} = \rho^{(n)} + \delta\rho \cdot \begin{cases} -1, & S_2 > \theta_2\, \rho < \rho_{\mathrm{sol}} \\ 0, & S_2 > \theta_2, \rho \geq \rho_{\mathrm{sol}} \\ +1/10\,, & S_2 \leq \theta_2 \end{cases}$$

with $\theta_2$ the threshold. The density is constrained to remain within $[\rho_{\mathrm{min}}, 1]$. Stage 2 runs for $N_2$ iterations.

The transition between stages is prescribed by $n_1$. This is a simplification: the actual timing in vivo is not known, and the two stages may overlap or alternate. We consider the sequential model as a first approximation and leave more complex staging protocols for future work.

### 2.4. Numerical implementation

The model is implemented in Abaqus 6.14 using user subroutines (FILM, USDFLD, UEXTERNALDB, URDFIL). The stationary diffusion problem is solved with the Abaqus solver using 8-node linear heat transfer brick elements (DC3D8). This element type is chosen because the model exploits the thermal analogy: the concentration field $u$ corresponds to temperature, and the diffusion coefficient $D$ corresponds to thermal conductivity. The density $\rho$ updates and boundary condition switching are handled by the user subroutines. Global data exchange is performed via SMA arrays. The flux $\mathbf{J}$ is treated as a generic matter flux throughout the implementation; its physical interpretation differs between stages but the numerical treatment is identical.

## 3. Results and Discussion

### 3.1. Reference calculation

We first present a reference calculation that illustrates the two-stage behavior of the model. The parameters are listed in Table 1. The computational domain is a quarter-model of a square active layer ($50 \times 50 \times 5$ elements), with a single-element seed at the symmetry center ($\rho = 1$). The diffusion problem is solved in the surrounding volume with the boundary conditions described in Section 2.1.

**Table 1.** Parameters of the reference calculation.

| Parameter | Symbol | Value |
|---|---|---|
| Diffusion coefficient | $D$ | 1 |
| Source concentration | $u_0$ | 100 |
| Sink concentration | $u_s$ | 10 |
| Film coefficient (active) | $h_{\mathrm{film}}$ | 5000 |
| Film coefficient (passive) | $h_{\mathrm{film}}$ | 0 |
| Minimum density | $\rho_{\min}$ | 0.01 |
| Maximum density | $\rho_{\max}$ | 1.0 |
| Activation threshold (Stage 1) | $\rho_{\mathrm{act}}^{(1)}$ | 0.7 |
| Activation threshold (Stage 2) | $\rho_{\mathrm{act}}^{(2)}$ | 0.2 |
| Dissolution threshold | $\rho_{\mathrm{sol}}$ | 0.7 |
| Stimulus threshold (Stage 1) | $\theta_1$ | 2800 |
| Stimulus threshold (Stage 2) | $\theta_2$ | 2600 |
| Regularization parameter | $\varepsilon$ | 0.001 |
| Number of iterations (Stage 1) | $n_1$ | 1900 |
| Total number of iterations | $n_{total}$ | 2500 |
| Density update step | $\delta\rho$ | 0.01 |

The evolution of the density field is shown in Fig. 3. At early stages ($n = 100$, panel a), the growth is uniform, without dendritic branching. The first branching event occurs at $n = 250$ (panel b). As the dendrites grow, they thicken near the domain boundary (panel e, $n = 1500$), consistent with the higher flux at the edges due to the additional flux from the boundary faces. By the end of Stage 1 (panel f, $n = 1900$), the dendritic pattern is fully developed.

In Stage 2, the morphology evolves in three sub-stages. First, the dendrites thicken (panel g, $n = 2000$). Then, bridges form between adjacent dendrites, creating pores (panel h, $n = 2100$). By $n = 2200$ (panel i), bridge and pore formation is complete. Finally, the density increases in the regions outside the pores (panel j, $n = 2500$), resulting in a regular network with approximately uniform pore spacing. Pores ($\rho < 0.4$) are shown in white.

The transition between stages is prescribed by $n = 1900$ iterations. The same visualization settings are used for all panels: density filter with radius $R = 1.4$, isosurface at $\rho = 0.4$. An animation of the full evolution is provided in the supplementary material.

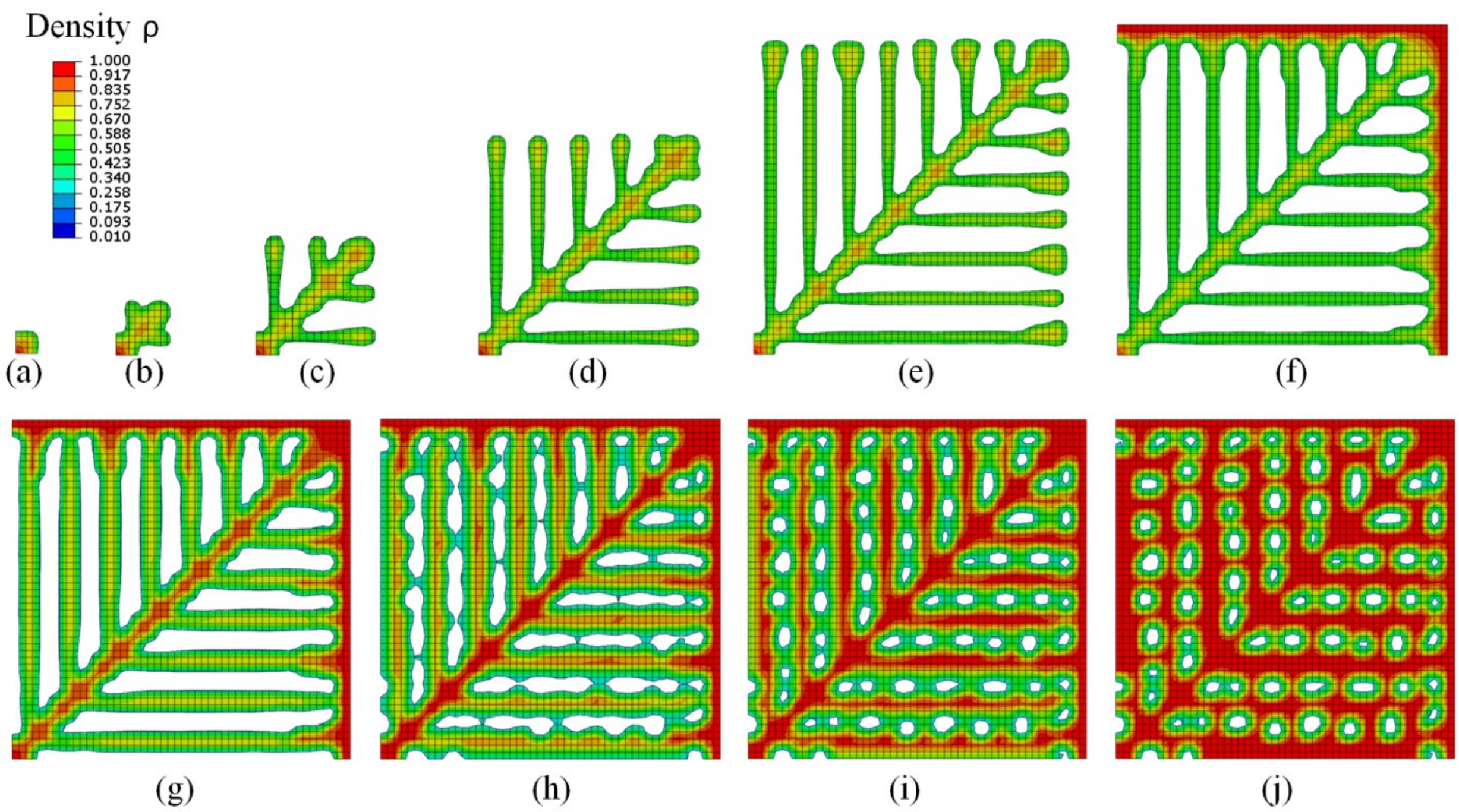

**Figure 3.** Evolution of the density field during the reference calculation. (a–f) Stage 1 (dendritic growth) at $n = 100, 250, 500, 1000, 1500, 1900$. (g–j) Stage 2 (pore and bridge formation) at $n = 2000, 2100, 2200, 2500$. White regions correspond to pores ($\rho < 0.4$). Density filter with radius $R = 1.4$ is applied for visualization. An animation of the full evolution is available at https://doi.org/10.5281/zenodo.23808711

### 3.2. Stage 1: morphology versus threshold

To analyze the effect of the threshold on the Stage 1 morphology, we performed a series of calculations with $\theta_1$ varying from 1400 to 3600 in steps of 200, keeping all other parameters fixed (Table 1). The results are shown in Fig. 4. The threshold controls the balance between the diffusive supply of matter and its consumption at the active layer. At high $\theta_1$, the stimulus $S_1 = |J|^2/\rho$ rarely exceeds the threshold, and growth is suppressed. At low $\theta_1$, the stimulus exceeds the threshold almost everywhere, and the layer fills uniformly. Between these limits, four distinct regimes appear.

**Regime I: suppressed growth** ($\theta_1 \geq 3400$, Fig. 4a,b). At $\theta_1 = 3600$ (Fig. 4a), the front does not advance: the initial seed remains isolated. At $\theta_1 = 3400$ (Fig. 4b), the front advances slightly and then arrests. In both cases, no dendritic structure develops.

**Regime II: dendritic growth** ($\theta_1 =$ 2800-3200, Fig. 4c–e). At $\theta_1 = 3200$ (Fig. 4c), the front becomes unstable and splits into separate branches. At $\theta_1 = 3000$ (Fig. 4d) and $\theta_1 = 2800$ (Fig. 4e), the number of dendrites increases. Notably, branching itself does not thin the branches: when a dendrite splits, each daughter branch has approximately the same thickness as the parent (see Discussion). This regime corresponds to the classical Mullins–Sekerka instability of a growth front [31].

**Regime III: thickening and merging** ($\theta_1 =$ 1800-2400, Fig. 4f–h). At $\theta_1 = 2400$ (Fig. 4f), the dendrites begin to thicken. At $\theta_1 = 2200$ (Fig. 4g), they continue to thicken and start to merge with adjacent branches. At $\theta_1 = 1800$ (Fig. 4h), merging becomes pronounced: adjacent branches coalesce, and the structure becomes denser.

**Regime IV: filling** ($\theta_1 \leq 1600$, Fig. 4i,j). At $\theta_1 = 1600$ (Fig. 4i), most dendrites have merged, leaving isolated pore regions. At $\theta_1 = 1400$ (Fig. 4j), the domain is completely filled.

The transitions between regimes are continuous. The number of dendrites first increases with decreasing $\theta_1$ (Regime II), then decreases as branches merge (Regime III). This non-monotonic behavior reflects the competition between nucleation of new branches and coalescence of existing ones.

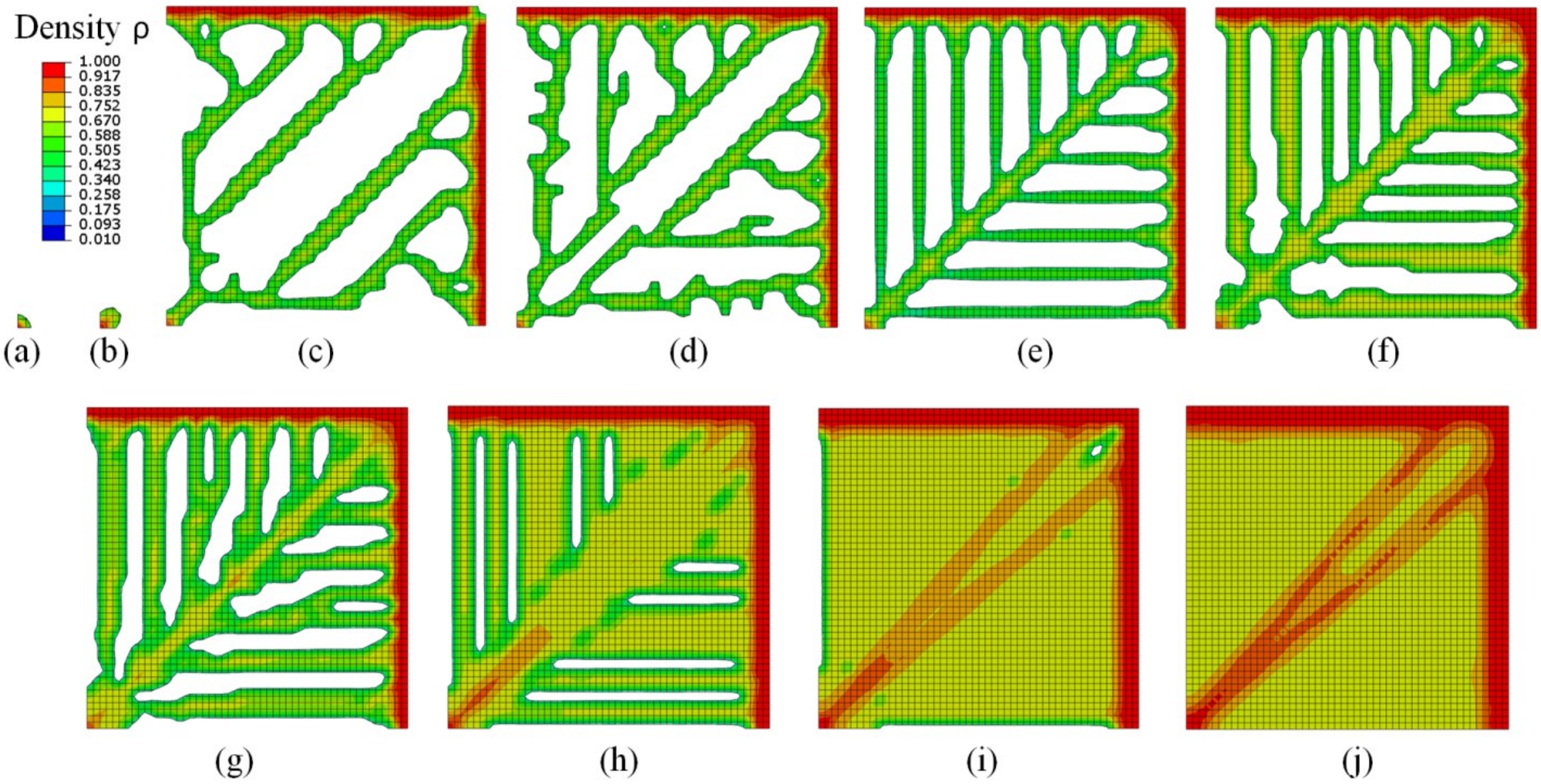

**Figure 4.** Stage 1 morphologies for varying $\theta_1$. (a) $\theta_1 = 3600$: no growth. (b) $\theta_1 = 3400$: arrested front. (c) $\theta_1 = 3200$, (d) $\theta_1 = 3000$, (e) $\theta_1 = 2800$: dendritic growth. (f) $\theta_1 = 2400$, (g) $\theta_1 = 2200$, (h) $\theta_1 = 1800$: dendrites thicken and merge. (i) $\theta_1 = 1600$: merged dendrites with isolated pores. (j) $\theta_1 = 1400$: full filling. White regions correspond to pores ($\rho < 0.4$).

### 3.3. Stage 2: morphology versus threshold

To analyze the effect of the threshold on the Stage 2 morphology, we performed a series of calculations with $\theta_2$ varying from 400 to 5000 in steps of 200, keeping all other parameters fixed (Table 1). The results are shown in Fig. 5.

In Stage 2, the update rule is inverted: growth occurs when the stimulus $S_2 = |J|^2/(1 - \rho + \varepsilon)$ is below the threshold, and decay occurs when it is above. The threshold thus controls the balance between pore formation (low $\theta_2$) and filling (high $\theta_2$). Four regimes appear.

**Regime I: no filling** ($\theta_2 \leq 400$, Fig. 4a). At low ($\theta_2$, the inter-dendritic space remains empty: no pores form.

**Regime II: elongated and isolated pores** ($\theta_2$ =400-1800, Fig. 4b). The inter-dendritic space becomes partially filled. Large elongated (merged) pores form together with isolated pores.

**Regime III: isolated pores only** ($\theta_2$ =2000-4800, Fig. 4c–g). As $\theta_2$ increases, the number of isolated pores decreases, and pores do not form between some dendrites. At $\theta_2 = 2600$ (Fig. 4d), the pore network is most regular. At $\theta_2 = 3600$ (Fig. 4f), pores form only between a few dendrites. At $\theta_2 = 4600$ (Fig. 4g), only isolated single pores remain.

**Regime IV: no pores** ($\theta_2 \geq 5000$, Fig. 4h). At high $\theta_2$, the inter-dendritic space is completely filled: no pores remain.

The transitions between regimes are continuous. Within Regime III, the number of pores decreases monotonically with increasing $\theta_2$.

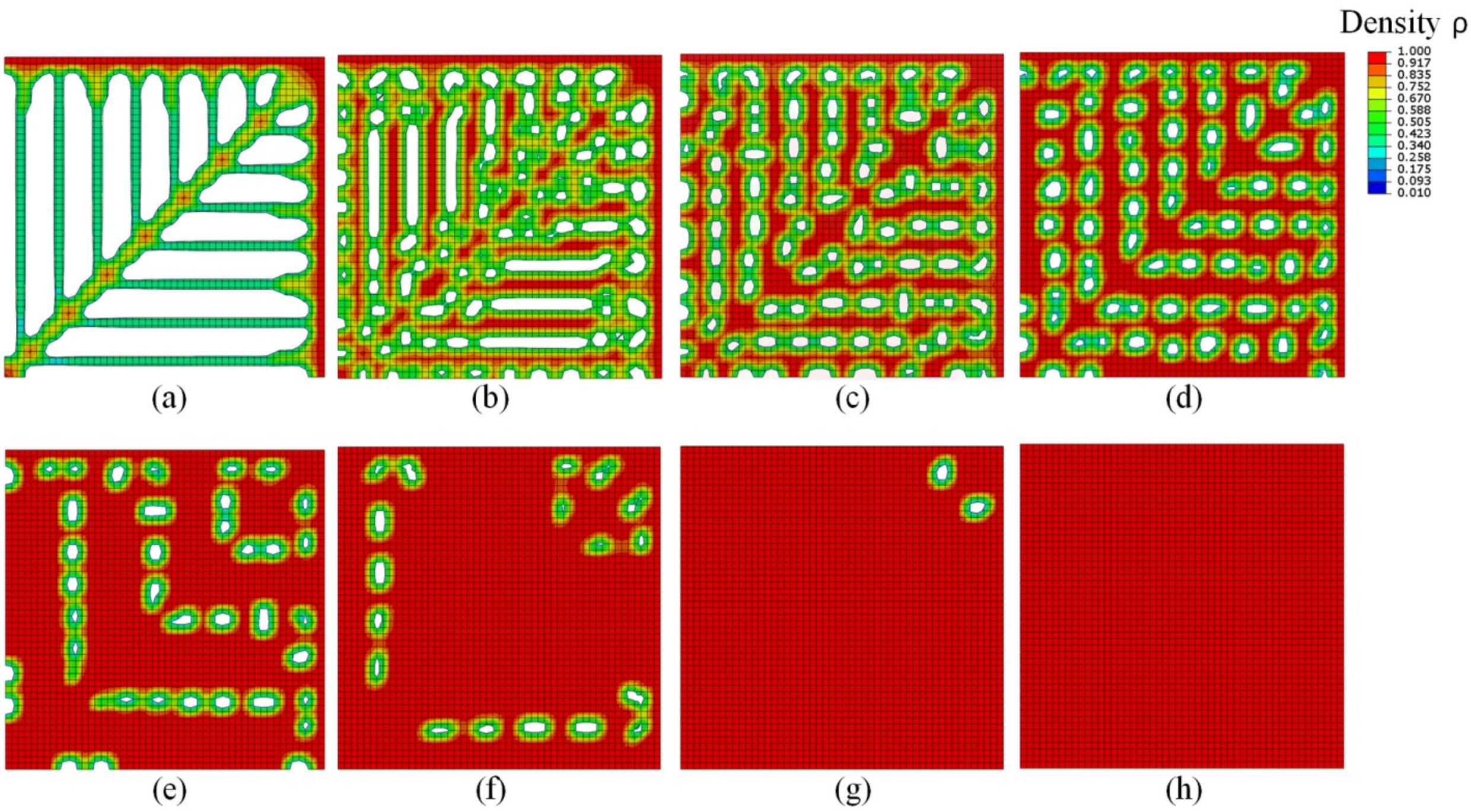

**Figure 5.** Stage 2 morphologies for varying $\theta_2$. (a) $\theta_2 = 400$: no filling. (b) $\theta_2 = 1400$: elongated and isolated pores. (c) $\theta_2 = 2000$, (d) $\theta_2 = 2600$, (e) $\theta_2 = 3200$, (f) $\theta_2 = 3600$, (g) $\theta_2 = 4600$: isolated pores only, with decreasing density. (h) $\theta_2 = 5000$: no pores. White regions correspond to pores ($\rho < 0.4$).

### 3.4. Quantitative characterization of morphologies

To quantify the morphologies described in Sections 3.2 and 3.3, we computed two characteristics for each threshold: the number of structural elements (dendrites in Stage 1, pores in Stage 2) and the volume fraction of material $\phi$. The results are shown in Fig. 6. The number of dendrites $N_1$ and pores $N_2$ were counted from the filtered density fields, and the volume fraction was computed as the average density over the active layer.

**Stage 1** (Fig. 6a). The number of dendrites $N_1$ (blue curve, left axis) is zero at $\theta_1 = 3380$ and increases with decreasing $\theta_1$, reaching a plateau of 15 at $\theta_1$ =2400-2800. This plateau corresponds to the regime of maximum branching, where dendrites have not yet begun to merge. At $\theta_1 < 2400$, $N_1$ decreases as branches merge.

The volume fraction $\phi_1$ (red curve, right axis) is zero at $\theta_1 = 3380$ and increases with decreasing $\theta_1$. Two distinct mechanisms contribute to this increase. First, between $\theta_1 = 3380$ and $3200$, the increase in $\phi_1$ is due to the onset of dendritic growth and the increasing number of dendrites. Second, between $\theta_1 = 2000$ and $1600$, the increase is due to the rapid merging of dendrites. Between these two intervals, $\phi_1$ increases more gradually.

**Stage 2** (Fig. 6b). The quantitative analysis of pores is performed starting from $\theta_2 = 2000$, which corresponds to the onset of Regime III (Section 3.3). At lower thresholds, the inter-dendritic space is filled with elongated and merged pores, making it difficult to distinguish individual pores. At $\theta_2 = 2000$, the number of individual pores $N_2$ (blue curve, left axis) is 92. As $\theta_2$ increases, $N_2$ decreases. The decrease is gradual between $\theta_2 = 2000$ and $2600$. A more rapid change is observed between $\theta_2 = 2800$ and $3600$, which is associated with the complete filling of the spaces between adjacent dendrites without pore formation. Because the number of dendrites is discrete, the number of pores changes in discrete steps. At $\theta_2 = 5000$, no pores remain. The volume fraction $\phi_2$ (red curve, right axis) starts from approximately 0.885 at $\theta_2 = 2000$ and reaches 1 at $\theta_2 = 5000$. The high initial value is due to the fact that Stage 2 begins after Stage 1, at which point the dendritic structure already occupies about 62% of the domain.

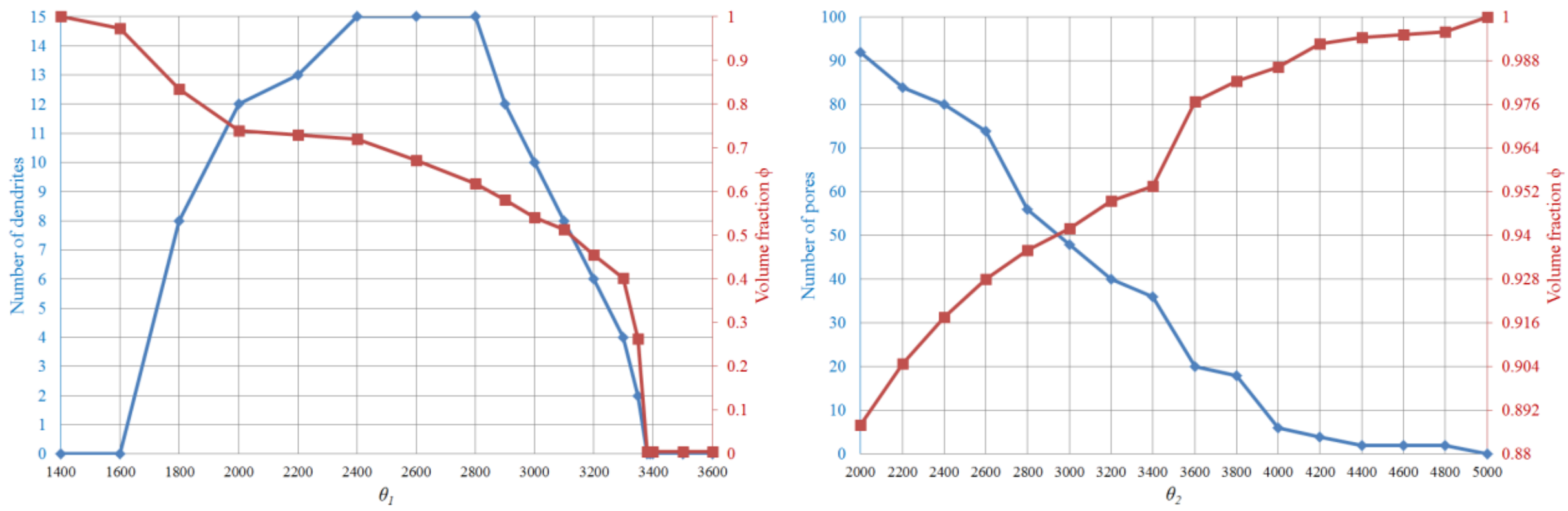

(a) (b)

**Figure 6.** Quantitative characterization of Stage 1 and Stage 2 morphologies.

(a) Stage 1: number of dendrites $N_1$ (blue, left axis) and volume fraction $\phi_1$ (red, right axis) versus $\theta_1$. (b) Stage 2: number of pores $N_2$ (blue, left axis) and volume fraction $\phi_2$ (red, right axis) versus $\theta_2$.

### 3.5. Discussion

The update rule used in the present model, based on the stimulus $S_1 = |J|^2/\rho$, is analogous to the sensitivity of compliance in density-based topology optimization [25], [26] and to the strain energy density per unit mass used in bone remodeling simulations [30]. This places the model within a unified framework of adaptive remodeling. However, the present model differs from classical SIMP in two important aspects. First, the diffusion coefficient $D$ does not depend on the density $\rho$ (Section 2.2). Second, the boundary condition is a fixed concentration on the source face, and the sink strength is dynamic, determined by the density field. As a result, the branch thickness in Stage 1 remains approximately constant during branching (Fig. 3c–e), whereas in classical topology optimization with uniform volumetric supply and fixed sinks, branches become thinner upon splitting because the available flux is divided among more branches. This setup is motivated by the geometry of the SDV, where silica is supplied through the membrane rather than uniformly from the volume.

Stage 1 reproduces the classical Mullins–Sekerka instability of a growth front [31]: a higher threshold suppresses branching, while a lower threshold promotes it (Section 3.2). Stage 2 uses an inverted update rule (Section 2.3). This inversion is motivated by the physical role of pores as transport channels. The high local flux through a pore suppresses further precipitation, stabilizing the pore. This creates a self-sustaining mechanism: a pore remains open because the flux through it is high, and the flux remains high because the pore is open. This behavior is conceptually similar to the functioning of pores in marine sponges, where flow actively maintains pore patency. The inversion of the update rules in Stage 2 was inspired by the classical topology optimization problem of a uniformly loaded block on a flat support, whose optimal solution consists of equally spaced columns. In our model, as a result of the inverted rules, these material columns turn into regularly spaced sink pores.

The two-stage sequence (dendritic ribs followed by pore formation) is consistent with recent cryo-ET observations of the silica deposition vesicle in *Thalassiosira pseudonana* [17]. In particular, the model reproduces the sequence of events: ribs grow and branch first, and pore formation occurs subsequently between the branches. This sequence is also consistent with classical ontogenetic studies of pennate diatoms, where a primary band and transapical ribs are formed first, followed by the fusion of spine-like projections to delimit the pores [32]. Figure 7 shows the full structure obtained by reflecting the quarter-model solution. The regular pore network and the absence of pores near the edges are visible. The edge effect is consistent with observations on diatom valves, where pores are typically absent in the marginal zone.

Figure 1 provides direct biological motivation for this two-stage sequence. Panel (a) shows the sequential valve development in *Navicula pelliculosa*: transapical ribs grow first, followed by the formation of cross-bridges that delimit pores. Panel (b) shows a partially dissolved valve of *Stephanodiscus hantzschii* with a regular hexagonal pore network. The present model reproduces both the sequence (Stage 1 → Stage 2) and the final regular pore network.

The model has several limitations. It is formulated in dimensionless units, and the parameters $\theta_1$ and $\theta_2$ are not directly linked to measurable biological quantities. The transition between stages is prescribed by a fixed number of iterations $n_1$; the actual timing in vivo is unknown, and the two stages may overlap or alternate. The finite element mesh introduces a slight anisotropy, which affects the direction of dendritic growth. Future work will focus on distinguishing the physical components of the flux, allowing the two stages to act simultaneously, and comparing the predicted pore statistics with experimental data.

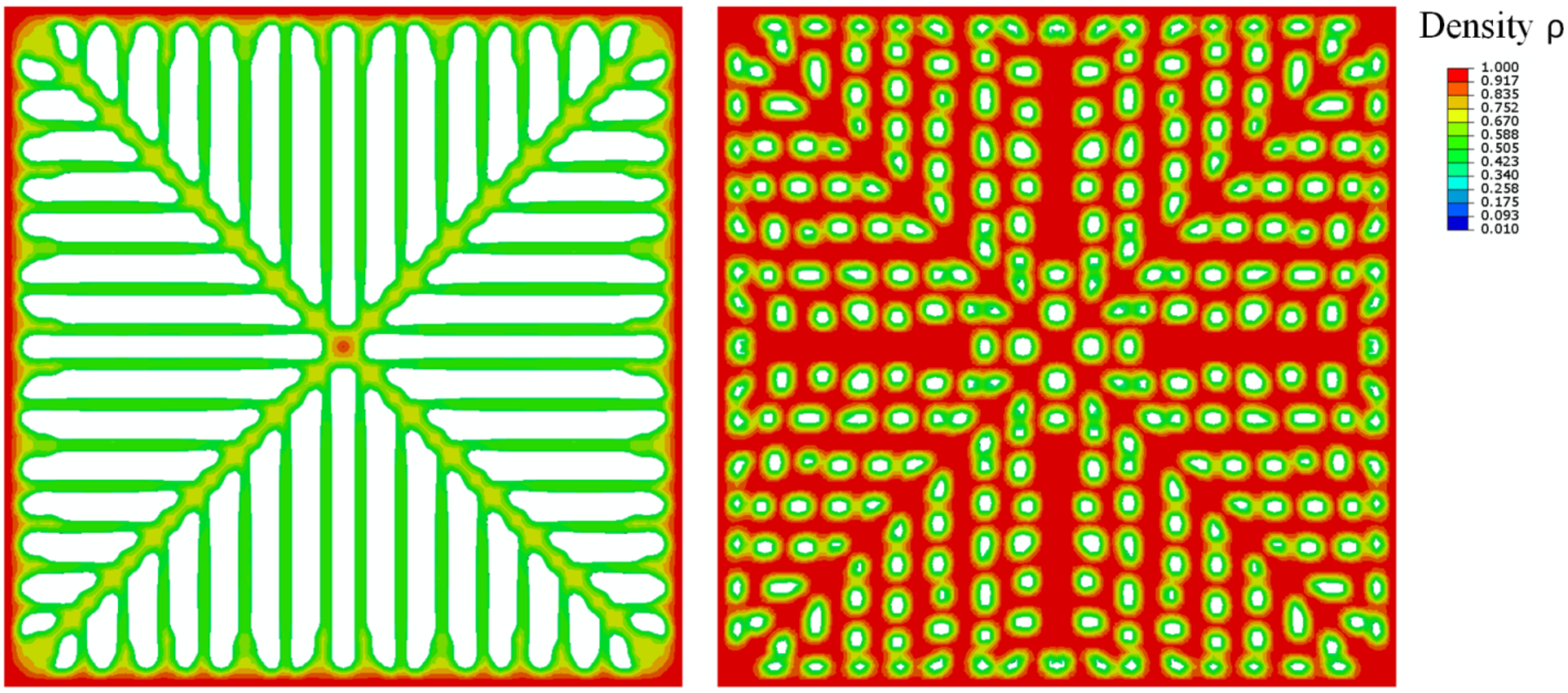

(a) (b)

**Figure 7.** Full structure obtained by reflecting the quarter-model solution across the symmetry axes. (a) End of Stage 1: dendritic pattern. (b) End of Stage 2: regular pore network. Pores are absent in the central region (initial seed) and near the domain edges (edge effect).

## 4. Conclusions

We have proposed a two-stage diffusion-driven model of morphogenesis with evolving connectivity. The model is formulated on a fixed finite element mesh and uses a density field that controls the boundary conditions for diffusion. The update rule in the first stage is analogous to sensitivity in SIMP topology optimization and to strain energy density in bone remodeling, while the inverted rule in the second stage produces regular pore networks. The model reproduces the observed sequence of morphological events: dendritic growth (Stage 1) followed by pore and bridge formation (Stage 2). It predicts four regimes in Stage 1 depending on the threshold $\theta_1$ and four regimes in Stage 2 depending on the threshold $\theta_2$, defining a phase space of morphologies. The predicted pore network is regular, with approximately uniform spacing, and pores are absent near the domain edges, consistent with observations in diatom valves.

The model has several limitations. It is formulated in dimensionless units, and the transition between stages is prescribed by a fixed number of iterations. Future work will focus on distinguishing the physical components of the flux, allowing the two stages to act simultaneously, and comparing the predicted pore statistics with experimental data.

## Acknowledgements

This research did not receive any specific grant from funding agencies in the public, commercial, or not-for-profit sectors.

## CRediT authorship contribution statement

**Alexander Safonov:** Conceptualization, Methodology, Software, Validation, Formal analysis, Investigation, Resources, Data curation, Writing – original draft, Writing – review & editing, Visualization, Project administration.